\documentclass[a4paper, amsfonts, amssymb, amsmath, nofootinbib, twoside, superscriptaddress,longbibliography]{revtex4-2}
\usepackage[english]{babel}
\usepackage[utf8]{inputenc}
\usepackage[colorinlistoftodos, color=green!40, prependcaption]{todonotes}
\usepackage{cancel,amsmath}
\usepackage{amsthm}
\usepackage{mathtools}
\usepackage{xcolor}
\usepackage{graphicx}
\usepackage{float}
\usepackage{soul}

\usepackage{gensymb}
\usepackage{multirow}
\usepackage[left=15mm,right=15mm,top=35mm,columnsep=15pt]{geometry} 
\usepackage{adjustbox}
\usepackage{placeins}
\usepackage[T1]{fontenc}

\usepackage{csquotes}

\usepackage[pdftex, pdftitle={Article}, pdfauthor={Author}]{hyperref}

\newcommand{\upv}{UPV}

\usepackage{algorithm2e}
\SetKwComment{Comment}{/* }{ */}
\RestyleAlgo{ruled}

\usepackage{cleveref}
\crefname{figure}{figure}{figures}
\Crefname{figure}{Figure}{Figures}
\crefname{section}{section}{sections}
\Crefname{section}{Section}{Sections}
\crefname{equation}{equation}{equations}
\Crefname{equation}{Equation}{Equations}
\crefname{algorithm}{algorithm}{algorithms}
\Crefname{algorithm}{Algorithm}{Algorithms}

\begin{document}
\title{Using Diffusion Models to Estimate Uncertainties in Analytic Continuation}

\author{Sagi Meir}
    \affiliation{School of Chemistry, Tel Aviv University, Tel Aviv 6997801, Israel.}
    \affiliation{The Center for Physics and Chemistry of Living Systems, Tel Aviv University, Tel Aviv 6997801, Israel.}
    
\author{Daniel Freedman}
    \affiliation{School of Mathematical Sciences, Tel Aviv University, Tel Aviv 6997801, Israel.}

\author{Barak Hirshberg\footnote{Author to whom any correspondence should be addressed hirshb@tauex.tau.ac.il}}
    \affiliation{School of Chemistry, Tel Aviv University, Tel Aviv 6997801, Israel.}
    \affiliation{The Center for Physics and Chemistry of Living Systems, Tel Aviv University, Tel Aviv 6997801, Israel.}
    \affiliation{The Center for Computational Molecular and Materials Science, Tel Aviv University, Tel Aviv 6997801, Israel.}

\begin{abstract}
    Inverse problems are ubiquitous in physics, chemistry, and engineering, arising when reconstructing hidden quantities from indirect measurements. A key example is the analytic continuation of imaginary-time correlation functions (iTCFs) to the real-frequency domain. This process requires an inverse Laplace transform, which is inherently ill-posed and highly sensitive to small input variations. Recent neural network (NN)-based methods have shown promising results by learning mappings from imaginary-time to real-frequency spectra, often outperforming traditional techniques such as maximum entropy. However, because the problem is ill-posed, many spectra fit the same iTCF. Regression-based approaches output a single solution, which approximates an average over the true solution space, and therefore fail to capture the full distribution of plausible power spectra. To address this issue, we introduce a diffusion-based framework for analytic continuation that learns the distribution of spectra consistent with a given iTCF. It offers two key advantages. First, it quantifies uncertainty directly from the learned distribution. Second, by analyzing the spread and structure of this distribution, we can quantitatively assess the intrinsic hardness of each inversion problem. We measure this hardness with a new metric, the uncertainty pseudo-volume. Applying the framework to an iTCF from a path-integral molecular dynamics simulation of liquid parahydrogen, we obtain the self-diffusion coefficient with an error bar and flag a secondary high-frequency peak as a possible spurious artifact. In contrast to previous attempts at uncertainty quantification, our generative approach rests on a concrete probabilistic basis, providing a more theoretically grounded measure of confidence in the reconstructed power spectra.
\end{abstract}


\maketitle

\section{Introduction}

Inferring underlying quantities from indirect observations is central to many areas of science and engineering~\cite{yaman2013survey, lesnic2021inverse, bingham2024inverse}. 
A prominent example in chemistry and physics is numerical analytic continuation, which aims to recover real-frequency spectra from quantum imaginary-time correlation functions (iTCFs)~\cite{silver1990maximum, rabani2002calculation, tripolt2019numerical, ying2022analytic, shao2023progress}. Such spectral recovery is crucial for extracting experimentally measurable quantities, including quantum transport coefficients and relaxation times, from equilibrium simulations~\cite{gallicchio1998application, rabani2002calculation, habershon2007quantum, gunnarsson2010analytical, wlazlowski2012shear, reymbaut2015maximum, tiihonen2018computation, hamann2020dynamic}. Recovering the spectrum, however, requires an inverse Laplace transform. The forward transform is a highly smoothing operator, so distinct spectral features are compressed into nearly indistinguishable curves in the imaginary-time domain. Consequently, the inversion is severely ill-conditioned and fundamentally non-unique. It amplifies noise and is highly sensitive to data quality~\cite{mcwhirter1978numerical, shi2023rethinking}. Any meaningful reconstruction must therefore confront the intrinsic ambiguity of the solution space.

Historically, a variety of methods have been employed to tackle this problem, with maximum entropy remaining the most widely used~\cite{bryan1990maximum, silver1990maximum, gubernatis1991quantum, jarrell1996bayesian, levy2017implementation, bergeron2016algorithms}. The maximum entropy method imposes solution stability through prior knowledge and often yields smooth, physical reconstructions~\cite{rabani2002calculation, habershon2007quantum}.
However, its success depends on reliable prior knowledge~\cite{bergeron2016algorithms, tripolt2019numerical, rothkopf2020bryan}. Even when such knowledge is available, the results are biased towards it and remain sensitive to noise in the iTCF~\cite{gunnarsson2010analyticalME, wang2022reconstructing, shi2023rethinking}. A recent analysis finds that improving the prior knowledge improves the reconstruction more than reducing the noise in the iTCF~\cite{chuna2026noiseless}.
Stochastic analytic continuation addresses this limitation by generating a collection of plausible spectra rather than a single ``best'' estimate~\cite{sandvik1998stochastic, sandvik2016constrained}, but often struggles to distinguish true physical features from noise-induced artifacts~\cite{shao2023progress}. Pad\'e and related methods can resolve sharp features when the input data are extremely clean, but their performance degrades rapidly in the presence of noise, often through unstable extrapolation or spurious poles~\cite{Beach2000Reliable, gunnarsson2010analytical, Schott2016Analytic, tripolt2019numerical, Huang2025Barycentric}. Nevanlinna-based approaches enforce the correct analytic structure and can outperform Pad\'e methods~\cite{fei2021nevanlinna, fei2021analytical, iskakov2024triqs}, but they do not remove the fundamental information loss of the forward transform~\cite{tripolt2019numerical, goulko2017numerical, rothkopf2022bayesian} and can also suffer from numerical instability~\cite{huang2024reconstructing}.

Recent machine learning approaches have attempted to address this difficulty~\cite{fournier2020artificial, yoon2018analytic, xie2021analytic, kades2020spectral, zhang2022training, wang2022reconstructing, zhao2026analytic} by training neural network (NN) models to learn mappings from iTCFs to their corresponding spectra, effectively acting as data-driven inverse operators. These regression-based methods often outperform classical techniques such as maximum entropy in accuracy and speed, particularly when the input data are noisy~\cite{fournier2020artificial, yoon2018analytic, kades2020spectral}. However, most existing machine learning approaches formulate analytic continuation as a regression problem and predict exactly one spectrum per input iTCF. They collapse the solution space into a single estimate, biased toward a conditional mean. Analytic continuation, however, is inherently non-unique: many spectra fit the same imaginary-time data within noise tolerance. These methods therefore do not faithfully capture the full uncertainty of the inversion. Even when uncertainty estimates are provided~\cite{zhang2022training, raghavan2024uncertainty, chen2025high}, they take the form of pointwise error bars around a single spectrum rather than a joint distribution over plausible spectra. Training an ensemble of models also typically underestimates the uncertainty~\cite{schweighofer2023quantification}.

Generative modeling offers a natural way to overcome this limitation. Unlike regression-based NN models, generative models learn the probability distribution of solutions, providing a representation of the inversion uncertainty. A variational autoencoder has been applied to spectral reconstruction in lattice quantum chromodynamics~\cite{chen2021machine}. However, it averages over the latent space to return a single most-probable spectrum, and its error bands come from bootstrapping over training epochs rather than from the distribution of admissible solutions. Among generative approaches, diffusion models are well-suited to analytic continuation because they faithfully represent complex, high-dimensional distributions~\cite{ho2020denoising, kawar2022denoising, chung2024diffusionposteriorsamplinggeneral, daras2024surveydiffusionmodelsinverse, feng2023score, wu2024principled}.
Despite their success in domains such as computer vision~\cite{rombach2022high, ramesh2022hierarchicaltextconditionalimagegeneration}, natural language processing~\cite{li2022diffusion, gong2023diffuseqsequencesequencetext}, and structural biology~\cite{watson2023novo, corso2023diffdockdiffusionstepstwists}, diffusion models have not previously been applied to analytic continuation.

In this work, we introduce the first diffusion-based generative framework for analytic continuation of iTCFs. Our approach delivers two key advantages. First, it quantifies uncertainty directly from the learned distribution. Second, by analyzing the spread and structure of this distribution, we can assess the intrinsic hardness of each inversion problem, which by itself offers a new way to understand when and why inversion becomes challenging. We quantify that hardness with a new metric, the uncertainty pseudo-volume, and demonstrate the framework on both synthetic spectra and an iTCF from a path-integral molecular dynamics simulation of liquid parahydrogen. Together, these contributions establish a theoretically grounded and practically powerful framework for tackling one of the most difficult problems in computational physics and chemistry.

\section{Problem Setting: Retrieving Quantum Dynamics}

Here we outline the physical origins of the analytic continuation problem in quantum dynamics, specifically within the context of path-integral simulations. The core algorithmic methodology follows in \cref{sec:Diffusion}. Although our motivating application is quantum dynamics, the framework introduced there is not restricted to this setting. It applies to any inverse problem in which the observable data are related to the target quantity through a strongly smoothing transform.

Many computational methods for quantum systems, including path-integral simulations and quantum Monte Carlo, can efficiently evaluate equilibrium expectation values of observables, such as $\left\langle\hat{A}\right\rangle = \operatorname{Tr}\left[\exp\left(-\beta \hat{H}\right)\hat{A}\right]/Z$, where $\beta$ is the inverse temperature, $\hat{H}$ is the Hamiltonian of the system, and $Z=\operatorname{Tr}\left[\exp\left(-\beta \hat{H}\right)\right]$ is the partition function.
In contrast, computing real-time dynamical quantities is substantially more difficult. A central example is the real-time correlation function $c(t)=\left\langle \hat{A}(0)\hat{A}(t)\right\rangle$, which determines the dynamical response of the system. Direct evaluation of such quantities is hindered by the dynamical sign problem~\cite{makri1991feynman, krilov1999real, habershon2007quantum}, which leads to exponentially slow convergence in many numerical approaches.

\begin{figure}
\begin{center}
\centerline{\includegraphics[width=0.5\linewidth]{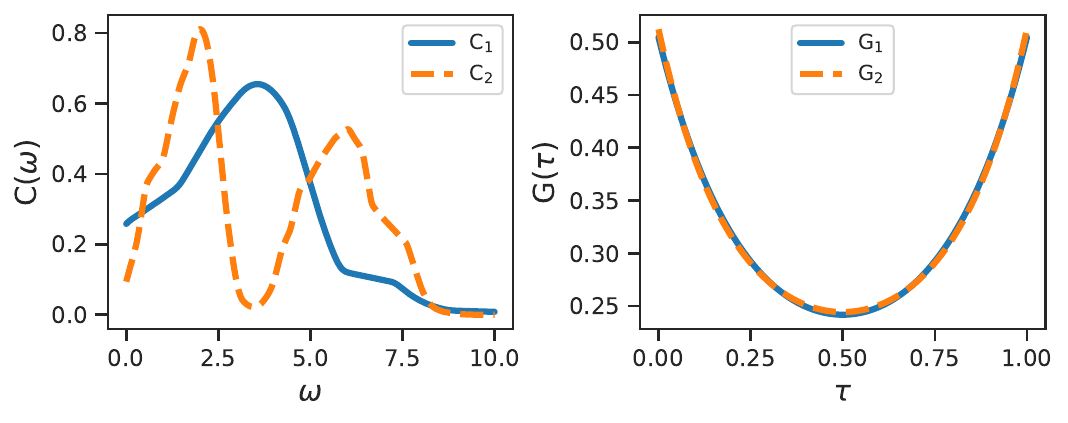}}
\caption{Demonstration of the ill-posedness of \cref{eq:G(tau)}. \textbf{Left panel:} Two synthetic spectra, $C_1$ (blue) and $C_2$ (orange), with clearly distinct features. \textbf{Right panel:} The corresponding iTCFs, $G_1$ and $G_2$, which are nearly indistinguishable.}
\label{fig:ill_posed_demo}
\end{center}
\end{figure}

A standard strategy for avoiding this difficulty is to work in imaginary time. Under the Wick rotation $t \mapsto -i\tau$, one obtains the iTCF $G(\tau)=\left\langle \hat{A}(0)\hat{A}(\tau)\right\rangle$ of $\hat{A}$, with $\tau \in [0,\beta\hbar]$, where $\hbar$ is the reduced Planck constant. Throughout this work, we use units in which $\beta\hbar=1$. 
Crucially, iTCFs converge numerically in a variety of equilibrium simulation techniques.
The experimentally accessible quantity, however, is the real-time dynamics, or equivalently the associated power spectrum $C(\omega)=\int_{-\infty}^{\infty}e^{i\omega t}c(t)\mathrm{d}t$. The two are related by the integral transform~\cite{rabani2002calculation},
\begin{equation}
    G(\tau) = 
    \frac{1}{2\pi}\int_{0}^{\infty}\left(e^{-\omega \tau}+e^{\omega (\tau-1)}\right)C(\omega)\mathrm{d}\omega.
\label{eq:G(tau)}
\end{equation}
Retrieving $C(\omega)$ from $G(\tau)$ is therefore essential for connecting finite-temperature simulations with experimentally measurable dynamical observables~\cite{gallicchio1998application, meyer2011transport, aarts2021electrical, rothkopf2022bayesian}. However, the inversion of \cref{eq:G(tau)} is essentially an inverse Laplace transform, which is severely ill-conditioned~\cite{shi2023rethinking}. This instability arises because the kernel $e^{-\omega \tau}+e^{\omega (\tau-1)}$ is strongly smoothing: fine spectral features in $C(\omega)$ are compressed into nearly indistinguishable imaginary-time signals. As a consequence, distinct spectra can generate almost identical iTCFs, and even small statistical or numerical errors in $G(\tau)$ can produce large variations in the reconstructed spectrum. The inverse problem is thus both unstable and non-unique (see demonstration in \cref{fig:ill_posed_demo}).

Machine learning methods have recently been proposed to approximate the inverse mapping from $G(\tau)$ to $C(\omega)$ by training NNs on synthetic data~\cite{fournier2020artificial, yoon2018analytic, xie2021analytic, kades2020spectral, wang2022reconstructing}. These approaches treat analytic continuation as a supervised regression problem, where the network outputs a single estimate of the spectrum for each input iTCF. While this strategy is more resilient to noise in the iTCF than classical regularization techniques, it introduces a fundamental limitation: the inversion is non-unique, yet regression models approximate only the conditional mean of the solution space. Consequently, such models cannot represent the full set of spectra compatible with a given $G$. That missing information is essential for uncertainty quantification and for assessing the intrinsic difficulty of the problem.

For this reason, we frame analytic continuation as a conditional generative modeling task. Rather than learning a deterministic inverse operator, we aim to model the conditional distribution of spectra given the observed iTCF. This probabilistic viewpoint preserves the multiplicity of admissible solutions and naturally supports uncertainty quantification. In the next section, we show how diffusion models provide a practical and theoretically grounded way to learn this distribution and to sample from it.

\section{Diffusion Models for Inverse Problems}
\label{sec:Diffusion}

Our goal is to model the conditional probability density $p(C \mid G)$ rather than to produce a single point estimate. Samples drawn from this distribution are then plausible spectral reconstructions conditioned on the observed data.
Generative models suit this setting because they are designed to model and sample from probability distributions.
Among generative modeling frameworks, we employ diffusion models because they represent broad and structured distributions while remaining stable to train~\cite{ho2020denoising, kawar2022denoising, daras2024surveydiffusionmodelsinverse}. Furthermore, recent works have shown that diffusion-based methods can be interpreted as approximate posterior samplers for noisy inverse problems, making them a natural choice when uncertainty quantification is a primary objective~\cite{chung2024diffusionposteriorsamplinggeneral, feng2023score, wu2024principled, song2023pseudoinverseguided}. 

\subsection{Conditional Diffusion Formulation}
\label{sec:diffusion_formulation}

Generative diffusion models approximate a target data distribution by learning to systematically invert a forward stochastic corruption process. Let the ground-truth power spectrum be denoted as $x_0 \equiv C(\omega)$, sampled from the underlying data distribution $p(x_0)$. Rather than employing the standard variance-preserving or variance-exploding stochastic differential equations~\cite{song2020score}, we adopt a continuous-time interpolant formulation akin to flow-matching~\cite{lipman2022flow}, with diffusion time $t \in [0,1]$.

In this framework, we define the forward process as a direct linear interpolation between the clean spectrum $x_0$ and standard Gaussian noise $\epsilon \sim \mathcal{N}(0, \mathbb{I})$, where $\mathbb{I}$ is the identity matrix. The corrupted state $x_t$ at time $t$ is then
\begin{equation}
    x_t = (1-t)x_0 + t\epsilon.
    \label{eq:forward_diffusion}
\end{equation}
This linear transformation yields a simple Gaussian conditional distribution for the corrupted data, $q(x_t | x_0) = \mathcal{N}(x_t; (1-t)x_0, t^2 \mathbb{I})$, which defines a probability path connecting the target data distribution at $t=0$ to a tractable noise prior at $t=1$. 

Inverting this process requires simulating the corresponding stochastic differential equation (SDE) backward in time~\cite{song2020score}. To achieve this, a model must learn the score function $\nabla_{x_t} \log q_t(x_t)$ of the perturbed data, where $q_t(x_t) = \int q(x_t \mid x_0)\, p(x_0)\, \mathrm{d}x_0$ is the marginal density at diffusion time $t$. The score is linearly related to the noise $\epsilon$ that generated the noisy sample. It is therefore common to learn either the score or the noise. We instead parameterize a Diffusion Transformer (DiT)~\cite{peebles2023scalable, chen2023pixartalphafasttrainingdiffusion} network $x_\theta(x_t, t, G)$ to directly estimate the clean data $x_0$ rather than the injected noise $\epsilon$. Because $\epsilon$ and $x_0$ are deterministically coupled via \cref{eq:forward_diffusion}, predicting $x_0$ is mathematically equivalent to predicting the score~\cite{song2021denoising, karras2022elucidating}. It provides, however, a substantially more stable optimization target in the continuous domain, particularly near $t=1$. Note that the network $x_\theta$ is a function of the noisy sample $x_t$, the time $t$, and the iTCF $G$.

The network is trained using a mean squared error objective:
\begin{equation}
    \mathcal{L}(\theta) = \mathbb{E}_{t, x_0, \epsilon} \left[ ||x_\theta(x_t, t, G) - x_0||^2 \right].
    \label{eq:loss}
\end{equation}
Optimizing this objective is closely related to minimizing the Kullback-Leibler divergence between the true posterior $p(C|G)$ and the learned model distribution~\cite{ho2020denoising, kingma2021variational}. During training, we sample the temporal variable $t$ from a Beta distribution, $t \sim \text{Beta}(\alpha, \beta)$, skewed toward zero. This non-uniform sampling reweights the loss landscape, forcing the network to allocate greater representational capacity to low-noise regimes, which is critical for resolving fine, highly correlated spectral features.

\subsection{Sampling and Inference}
\label{sec:diffusion_inference}

At inference, we generate spectral reconstructions by drawing samples from the learned conditional distribution $p(C|G)$. We initialize each trajectory from the prior, $x_{t_0} \sim \mathcal{N}(0, \mathbb{I})$, corresponding to a state of maximal entropy at $t_0 \approx 1$. We then numerically integrate the reverse-time process over a monotonically decreasing sequence of discrete timesteps $t_0 > t_1 > \dots > t_N \approx 0$.  While the reverse-time process is an SDE, it is always possible to convert an SDE to a corresponding ordinary differential equation (ODE)~\cite{song2020score}, which has a variety of numerical benefits. In particular, we gain efficiency because we may take longer timesteps. We pursue this direction below.

Because the neural network directly predicts the clean spectrum at each iteration, $\hat{x}_0^{(i)} = x_\theta(x_{t_i}, t_i, G)$, the sampling trajectory reduces to a sequence of deterministic updates. A standard first-order discretization of the reverse ODE yields the Denoising Diffusion Implicit Models (DDIM)~\cite{song2021denoising} update step. For our interpolant formulation, this first-order step is
\begin{equation}
    x_{t_{i+1}} = \frac{1}{t_i} \left( (t_i - t_{i+1})\hat{x}_0^{(i)} + t_{i+1} x_{t_i} \right).
    \label{eq:ddim_step}
\end{equation}

While \cref{eq:ddim_step} is sufficient for generation, first-order solvers typically require many function evaluations to minimize discretization errors. To accelerate convergence while maintaining high sample fidelity, we employ a second-order exponential integrator based on DPM-Solver++~\cite{lu2025dpm}. After taking an initial DDIM step for $i=0$, the algorithm uses the previous prediction $\hat{x}_0^{(i-1)}$ to construct a higher-order extrapolated update for all subsequent steps ($i \ge 1$),
\begin{equation}
    x_{t_{i+1}} = \frac{1}{t_i} \left( (t_i - t_{i+1})\hat{D}_i + t_{i+1} x_{t_i} \right),
    \label{eq:dpm_step1}
\end{equation}
where the effective higher-order data prediction $\hat{D}_i$ is
\begin{equation}
    \hat{D}_i = \left( 1 + \frac{h_{i+1}}{2h_i} \right)\hat{x}_0^{(i)} - \frac{h_{i+1}}{2h_i}\hat{x}_0^{(i-1)}.
    \label{eq:dpm_step2}
\end{equation}
Here the step sizes are $h_i = \log \left( \frac{1-t_i}{t_i} \right) - \log \left( \frac{1-t_{i-1}}{t_{i-1}} \right)$, which reflect the log-signal-to-noise ratio of the linear noise schedule.
A single integration of this ODE yields one spectral reconstruction. By generating an ensemble of independent trajectories, we map the posterior $p(C|G)$ and extract both the mean spectral prediction and the correlated uncertainty bounds of the inverse problem. \Cref{si:algorithms} gives pseudocode for training and sampling, and \cref{si:inference} lists the numerical settings we used.

\subsection{Overall Strategy, Model Architecture and Dataset}

Our model employs a Transformer-based architecture conditioned on the imaginary-time input $G(\tau)$. This choice is motivated by the non-local structure of the forward transform in \cref{eq:G(tau)}: each value of $G(\tau)$ contains integrated information from the full frequency range of $C(\omega)$, so reconstructing $C(\omega)$ depends on correlations across the entire temporal and spectral grids. An attention-based architecture combined with positional encoding is therefore well-suited to capture such long-range dependencies, overcoming the locality of standard convolutional networks.

The training strategy for the diffusion model follows the same general philosophy as earlier supervised regression approaches to analytic continuation, which use synthetic pairs of spectra and iTCFs to encode prior structural information~\cite{fournier2020artificial, yoon2018analytic, kades2020spectral, zhang2022training}. The key distinction is that we generate the synthetic data on-the-fly, making data generation itself a central component of the training method. Because the training dataset is not fixed in advance, the model is less prone to memorization and is exposed to a broader range of spectral shapes during optimization. Specifically, the random physical spectral functions $C(\omega)$ are generated by (i) choosing an initial function which is a mixture of Gaussians; (ii) modifying this initial function by smoothly warping the $\omega$-domain.   We inject randomness through the choice of the Gaussians and their weights, and through the warp, which we produce as a spline function with randomly chosen control points.  We then compute the corresponding iTCFs by forward trapezoidal integration of \cref{eq:G(tau)}.

\begin{figure}
\begin{center}
\centerline{\includegraphics[width=0.95\linewidth]{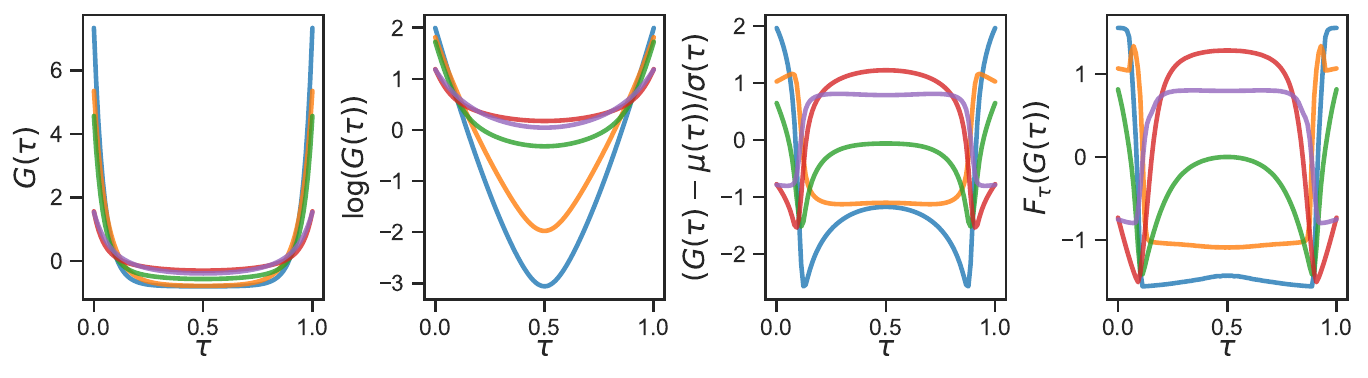}}
\caption{The four-channel representation of the iTCFs used for model conditioning. To provide a richer representation of the input signal and expose hidden structural features, the original iTCF is transformed into four complementary channels. The different colors represent a selection of sample iTCFs from the dataset. From left to right: The raw iTCF $G(\tau)$, which often compresses distinct spectral features into visually similar curves. The logarithmic transformation $\log(G(\tau))$, which enhances the visibility of low-amplitude structure, particularly near $\tau = 0.5$. The pointwise normalized signal $(G(\tau)-\mu(\tau))/\sigma(\tau)$, computed using the pointwise training-set mean and mean absolute deviation, which deemphasizes absolute scale to highlight relative variations. The pointwise histogram equalization $F_{\tau}(G(\tau))$, which maps the input through the empirical cumulative distribution function at each time slice to further distinguish overlapping features.}
\label{fig:G_four_channel}
\end{center}
\end{figure}

Our second methodological contribution is the representation of the conditioning iTCF. Instead of conditioning the diffusion model on a single channel containing $G(\tau)$, we use a four-channel representation that provides complementary views of the same signal, see \cref{fig:G_four_channel}. The first channel is the original iTCF $G(\tau)$. The second is $\log G(\tau)$, which enhances low-amplitude structure. The third is a pointwise normalized version of $G(\tau)$ defined using a precomputed pointwise training-set mean $\mu(\tau) = \frac{1}{N} \sum_{i=1}^N G^{(i)}(\tau)$ and pointwise mean absolute deviation $\sigma(\tau) = \frac{1}{N} \sum_{i=1}^N \left|G^{(i)}(\tau) - \mu(\tau)\right|$, computed over $N=200{,}000$ training samples $\{ G^{(i)}(\cdot)\}_{i=1}^N$. Each sample is then transformed as $\left(G^{(i)}(\tau)-\mu(\tau)\right)/\sigma(\tau)$, which deemphasizes absolute scale and highlights relative variation. The fourth uses pointwise histogram equalization. For each $\tau$, we construct the empirical cumulative distribution function $F_{\tau}$ of $G(\tau)$ over the training set and remap each sample as $F_{\tau}\left(G^{(i)}(\tau)\right)$. Finally, we apply a global channel-wise normalization so that all four channels reach the model on comparable scales. Together, these four channels provide a richer conditioning representation and enable the model to detect informative structure that may be less apparent in the raw iTCF alone. We give the full conditioning pipeline in \cref{si:conditioning}.
Additional implementation details, including architectural choices, conditioning strategy, and data-generation procedures, are provided in \cref{sec:comp_details}.

\section{Results and Discussion}

\subsection{Distributional Predictions}

\begin{figure}
\begin{center}
\centerline{\includegraphics[width=0.95\linewidth]{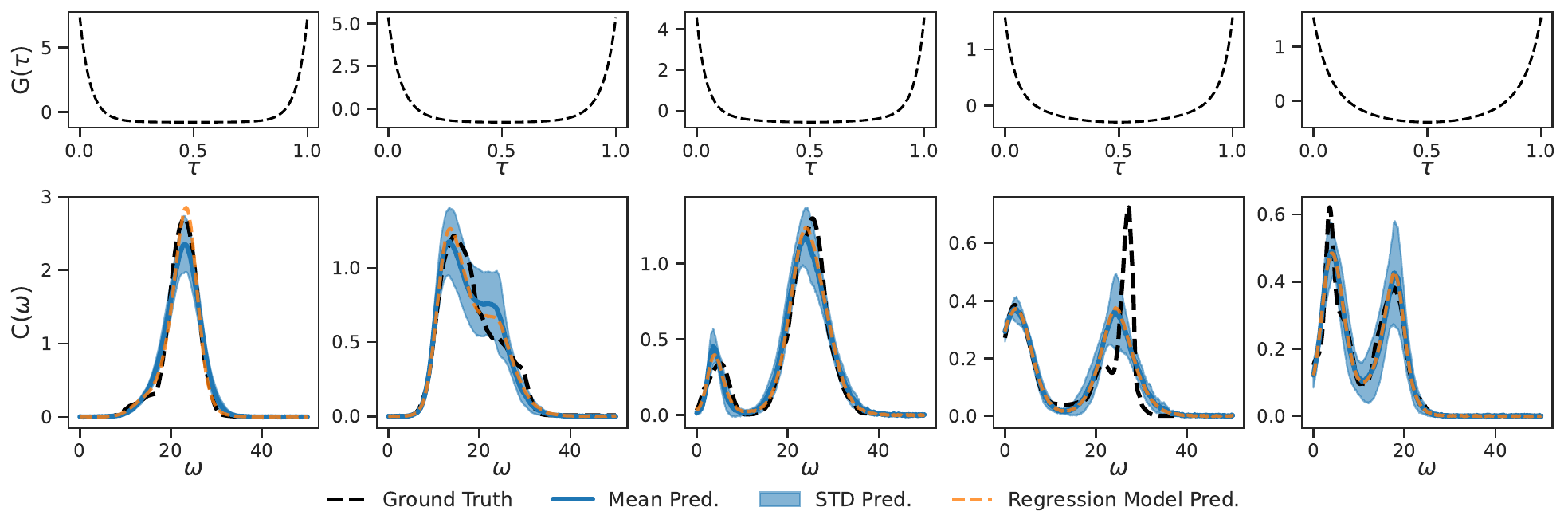}}
\caption{Comparison of analytic continuation using regression and diffusion models across representative synthetic examples. \textbf{Top row:} Ground-truth iTCFs obtained via \cref{eq:G(tau)} from the generated synthetic spectra. \textbf{Bottom row:} Ground-truth spectra (dashed black) compared against reconstructed spectral functions. The solid blue line and shaded light blue area are the mean and standard deviation of $1000$ realizations of the diffusion model. The prediction of the baseline regression model is shown in dashed orange.}
\label{fig:example_of_ac}
\end{center}
\end{figure}

The primary advantage of the diffusion framework is its ability to explicitly represent and quantify the intrinsic ambiguity of the analytic continuation problem. \Cref{fig:example_of_ac} illustrates this capability across representative synthetic examples. The top row displays the ground-truth imaginary-time correlation functions, $G(\tau)$.
Despite originating from spectrally distinct ground-truth signals, these iTCFs are visually similar, underscoring the fundamental non-uniqueness of the inversion.

The bottom row compares the ground-truth real-frequency spectra $C(\omega)$ (dashed black lines) against the model predictions.  The results for the diffusion model are displayed in blue: the solid blue curve in \cref{fig:example_of_ac} represents the mean of $1000$ realizations conditioned on the same $G(\tau)$, and the light blue shaded area represents the standard deviation.  For comparison, we trained a regression baseline employing the same Transformer architecture as the diffusion model, shown in dashed orange. While the regression model accurately approximates the conditional mean of the solution space (the solid blue and dashed orange curves are very close), it suppresses alternative admissible spectra for each input iTCF. In contrast, the diffusion model produces a distributional output using a single trained network, as illustrated by the shaded blue regions.

A common strategy for estimating uncertainty in regression-based approaches is to train an ensemble of independently initialized models on different datasets~\cite{fort2020deepensembleslosslandscape, ovadia2019can}. To evaluate this approach fairly, we trained multiple regression models on distinct on-the-fly generated datasets.  In practice, all models converged to nearly identical spectral reconstructions, indicating that the ensemble strategy is ineffective for capturing the solution variance.  In contrast, the variance of the diffusion model (illustrated by the shaded light blue region) is not a result of model instability or training noise, but reflects the uncertainty imposed by the kernel in \cref{eq:G(tau)}.

Finally, the bottom row of \cref{fig:ambiguity_metrics} shows example spectra generated for each iTCF of \cref{fig:example_of_ac}. Note the wide array of possible spectra that solve the inverse problem for a given iTCF. We analyze this quantitatively in \cref{sec:inversion_ambiguity}.

\subsection{Correlated Uncertainty and Parsimony of the Solution Space}
\label{sec:correlated}
While the variance shown in the shaded regions of \cref{fig:example_of_ac} provides an intuitive visual measure of uncertainty, it neglects the strong correlations between different frequencies imposed by the integral transform in \cref{eq:G(tau)}. As recently demonstrated in the context of imaging inverse problems~\cite{belhasin2023principal}, pointwise uncertainty estimates implicitly assume independence across coordinates (in our case, frequencies), leading to a gross overestimation of the effective uncertainty volume. As a result, such estimates fail to capture the true, highly structured geometry of the posterior distribution.

To obtain a tighter and more informative characterization of uncertainty, we construct a correlated uncertainty model, $p_p$, by performing principal component analysis (PCA) on the ensemble of spectra generated by the diffusion model.  Note that this is a \textit{local}, or iTCF-dependent, PCA: each iTCF $G$ will generate its own set of spectra from the diffusion model conditioned on $G$, and the PCA components are therefore specialized to that $G$. Another way to think of this is as a locally Gaussian approximation to the diffusion model $p(\cdot|G)$, where the parameters of the Gaussian depend on $G$. We show representative components in \cref{si:pca}.

We demonstrate that $p_p$ outperforms an uncorrelated model $p_u$ by assessing the efficiency of these basis representations using a parsimony analysis.  Specifically, let $\hat{C}_k$ be the best fit to $C$ using $k$ basis components.  We measure the efficiency of the representation by the normalized residual error $r_k$
\begin{equation}
    r_k = \frac{\left\|C-\hat{C}_k\right\|}{\|C\|},
\end{equation}
and tracking how it varies as $k$ increases. Here, $\|\cdot\|$ is the $L^2$-norm.

Let $V_k\in\mathbb{R}^{d\times k}$ be a matrix whose columns are the first $k$ principal components and $d$ is the dimensionality of the discretized spectrum.  Then for the correlated model $p_p$, the best fit can be shown to be $\hat{C}_k = \Pi_kC + (I-\Pi_k) \mu$ where the projection matrix $\Pi_k = V_k V_k^T$. The normalized residual error is then
\begin{equation}
    r_k^{(p)} = \frac{\sqrt{\|C-\mu\|^2 - (C-\mu)V_k V_k^T (C-\mu)}}{\|C\|}.
\end{equation}
A similar result can be used to compute $r_k^{(u)}$ for the uncorrelated model $p_u$. In this case, we can construct $V_k$ as follows: we sort the per-frequency standard deviations in descending order, and take $V_k$ to be the collection of standard basis vectors (i.e.~vectors which are all zeros except for a single 1), corresponding to the first $k$ such sorted frequencies.  However, to provide a more rigorous baseline, we grant the uncorrelated model $p_u$ a strict ``best-case advantage''. Instead of selecting this fixed $V_k$, we explicitly hand-pick the $k$ points with the largest absolute deviations from the mean \textit{for each sample}.  (This is in essence cheating, by choosing the best $V_k$ for a given sample.)  The corresponding residual error may therefore be written as 
\begin{equation}
    r_k^{(u)} = \frac{\sqrt{\|C-\mu\|^2 - \sum_{i \in \mathcal{I}_k} \left(C(\omega_i) - \mu(\omega_i)\right)^2}}{\|C\|},
\end{equation}
where $\mathcal{I}_k$ is the set of indices corresponding to the $k$ largest absolute deviations from the mean.

\begin{figure}
\begin{center}
\centerline{\includegraphics[width=0.99\linewidth]{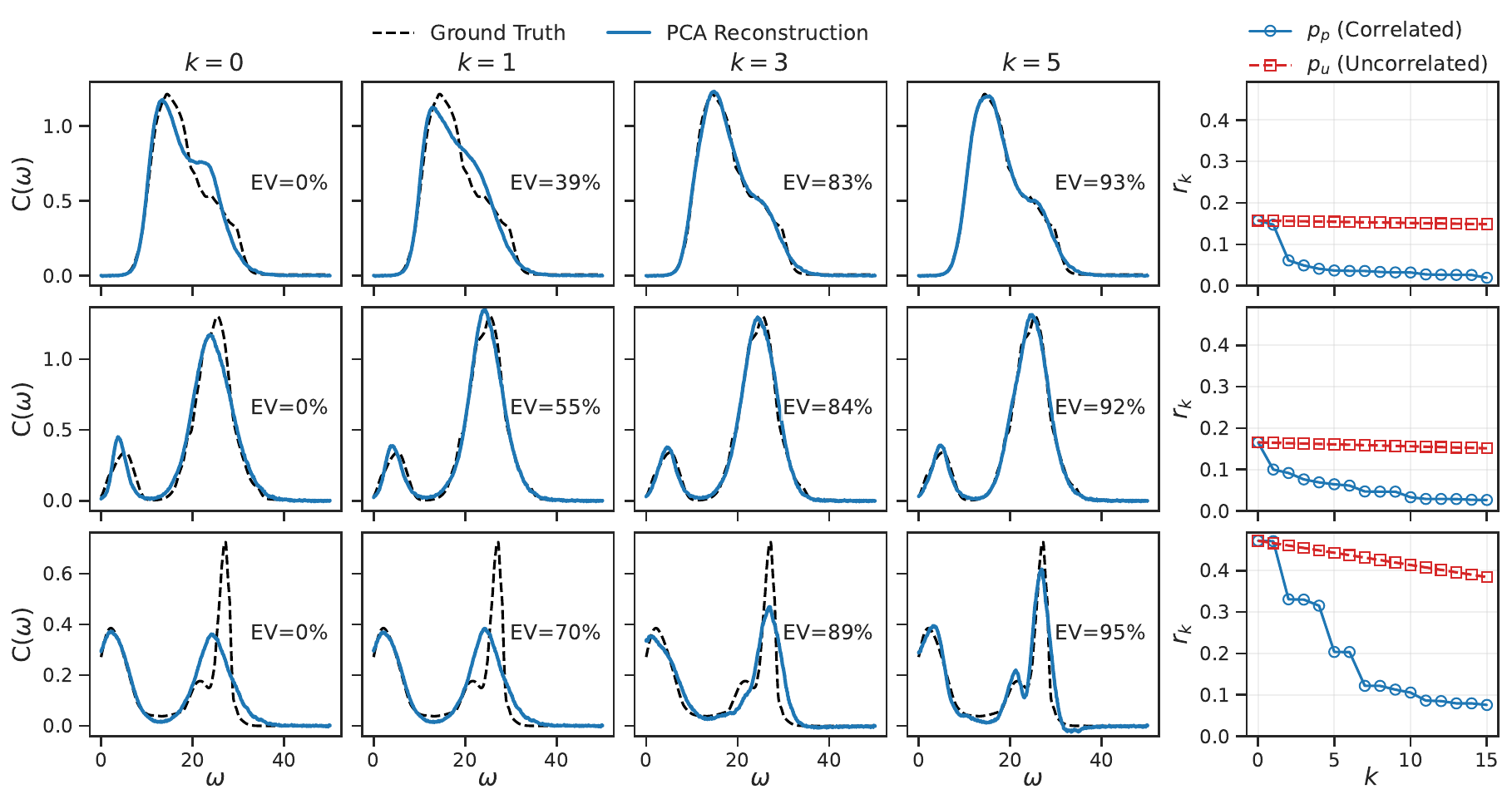}}
\caption{Step-by-step visualization of the PCA reconstruction for the correlated model ($p_p$). The \textbf{rightmost column} shows the parsimony analysis for both the correlated and uncorrelated models. Rows are three representative synthetic examples; columns show the number of principal components $k$ increasing from $0$ to $5$. Ground-truth spectra $C(\omega)$ (dashed black lines) are compared against the PCA reconstructions $\hat{C}_k(\omega)$ (solid blue lines). Annotations within each panel denote the cumulative explained variance (EV). The progression confirms the rapid convergence of the correlated model, with accurate reconstructions by $k=3$ to $k=5$. \textbf{Rightmost column:} Parsimony analysis evaluating the efficiency of the basis representation. The normalized residual error $r_k = \|C - \hat{C}_k\| / \|C\|$ is shown as a function of the number of components $k$ used in the reconstruction. For the correlated model ($p_p$, blue circles), we use the first $k$ principal components learned from the data.
For the uncorrelated model ($p_u$, red squares), we grant it a best-case advantage: instead of using fixed or random points, we explicitly select the $k$ grid points with the largest deviation from the mean ($|C(\omega_i) - \mu_i|$). Despite this advantage, the uncorrelated model converges slowly, whereas $p_p$ achieves superior accuracy with far fewer components ($k \approx 3$ versus $k \gg 10$).}
\label{fig:vis_converge}
\end{center}
\end{figure}

Despite this highly favorable construction, the uncorrelated model $p_u$ converges slowly as shown in the rightmost column of \cref{fig:vis_converge} (red squares). Specifically, one requires $k$ much larger than $10$ to achieve meaningful accuracy. In contrast, the correlated model $p_p$ (blue circles) exhibits rapid error decay and reaches near-convergence with only $k \approx 3$ components. This gap shows that accounting for cross-frequency correlations dramatically reduces the effective dimensionality of the solution space, consistent with sparse-dictionary formulations of the inversion~\cite{chuna2026discoveringwellconditionedanalyticcontinuation}. Therefore, the leading principal components capture the dominant correlated modes of variation in the learned conditional distribution.

\Cref{fig:vis_converge} provides a qualitative visualization of this rapid convergence. Each row corresponds to a representative spectrum, while columns show the correlated-model reconstructions $\hat{C}_k(\omega)$ as the number of retained components increases from $k=0$ to $k=5$. Starting from the dataset mean ($k=0$), each subsequent component resolves finer correlated spectral features. By $k=5$, the reconstructions align with the ground truth. At this stage, the cumulative explained variance typically reaches above 90\%, and the residual error $r_k$ falls below 10\%.

\subsection{Quantitative Analysis of Inversion Ambiguity}
\label{sec:inversion_ambiguity}
To move beyond qualitative assessment, we introduce a quantitative metric, the Uncertainty Pseudo-Volume (\upv{}), that characterizes both the magnitude and structure of inversion uncertainty. The idea is to find a hyper-ellipsoid which captures most of the variation of the spectra for a given iTCF, and then to use a volume-like measure on this hyper-ellipsoid to quantify the uncertainty.  In this sense, we follow a similar path to that introduced by \citet{belhasin2023principal}, but our measure is slightly different.

We begin by constructing the relevant hyper-ellipsoid using a PCA decomposition. This is the same PCA used for the correlated model $p_p$ in \cref{sec:correlated}, applied here to samples normalized by their average $L^2$-norm so that the measure is comparable across spectra of different scale. Specifically, for a given iTCF $G$ consider the set of $N$ diffusion-generated samples $\left\{C^{(j)} \right\}_{j=1}^N$ where $C^{(j)} \sim p(C|G)$.  We first normalize the samples by their average $L^2$-norm, i.e., $\bar{C}^{(j)}=C^{(j)}/\left( {\frac{1}{N} \sum_{j=1}^N\left\|C^{(j)} \right\|} \right)$. We then compute the set of principal components $\{v_i\}_{i=1}^d$ for this normalized set $\{\bar{C}^{(j)} \}$, where $d$ is the dimensionality of the spectra.  Given these principal components, we define the set of projection coefficients along the $i$-th principal component $v_i$, 
\begin{equation}
    S_i = \left\{ v_i^{\top}\left(\bar{C}^{(j)} - \mu \right): j \in \{1, \dots, N \} \right\}
    \quad \text{for } i = 1, \dots d
\end{equation}
where $\mu$ is the mean of $\{\bar{C}^{(j)} \}$. For each set $S_i$, we denote its 5th and 95th percentiles with $\ell_i$ and $u_i$, respectively, and define their difference $z_i\equiv u_i-\ell_i$.  The set $\{ z_i \}_{i=1}^d$ represents the axis lengths of the hyper-ellipsoid.

Next, we introduce a pseudo-volume, a measure defined on the axis lengths of the hyper-ellipsoid $\{ z_i \}_{i=1}^d$. There is a family of measures $\{\mathcal{P}_d\}$, one for each dimensionality $d$, which must satisfy three properties:
\begin{enumerate}
    \item $\mathcal{P}_d(\mathbf{0}) = 0$.
    \item $\frac{\partial \mathcal{P}_d}{\partial z_i} > 0$ for all $i = 1, \dots, d$.
    \item For any $z_1, \dots, z_d$ we have that $\mathcal{P}_d(z_1, \dots, z_{i-1}, 0, z_{i+1}, \dots, z_d) =  \mathcal{P}_{d-1}(z_1, \dots, z_{i-1}, z_{i+1}, \dots, z_d)$.
\end{enumerate}
Properties 1 and 2 are basic: if all axes have length zero, the pseudo-volume is correspondingly zero, and if any axis increases, then the pseudo-volume also increases.  Property 3 is more subtle.  It says that the pseudo-volume for a collection of $d$ axes, one of which is equal to zero, is the same as the pseudo-volume for $d-1$ axes with the zero axis dropped out.  We impose it because, even though we live in $d$ dimensions, the trailing axes of typical ellipsoids are nearly zero in size.  Property 3 therefore lets us treat the ellipsoid either as living in the full $d$-dimensional space, or, after removing the small eigenvalues (axes), as living in a $k$-dimensional space with $k < d$.  Both give nearly the same pseudo-volume.

It can be shown that the following family of functions are pseudo-volumes:
\begin{equation*}
    \mathcal{P}_d(z_1, \dots, z_d) = K \phi^{-1} \left( \sum_{i=1}^d \phi(z_i) \right)
\end{equation*}
where $K > 0$ and the function $\phi:\mathbb{R}_+ \to \mathbb{R}$ satisfies (1) $\phi(0) = 0$ (2) $\phi'(a) > 0$.  We choose $\phi(a) = \log(1 + \tfrac{a}{\lambda})$ for a positive constant $\lambda$. Taking $K = \tfrac{1}{\lambda}$ then gives the \upv{}
\begin{equation}
    \mathcal{P}_d(z_1, \dots, z_d) = \prod_{i=1}^d\left(1+\frac{z_i}{\lambda}\right)-1 
\label{eq:upv}
\end{equation}
Examining \cref{eq:upv}, we see that the \upv{} is quite similar to a standard volume measure, which is desirable.

\begin{figure}
\begin{center}
\centerline{\includegraphics[width=0.95\linewidth]{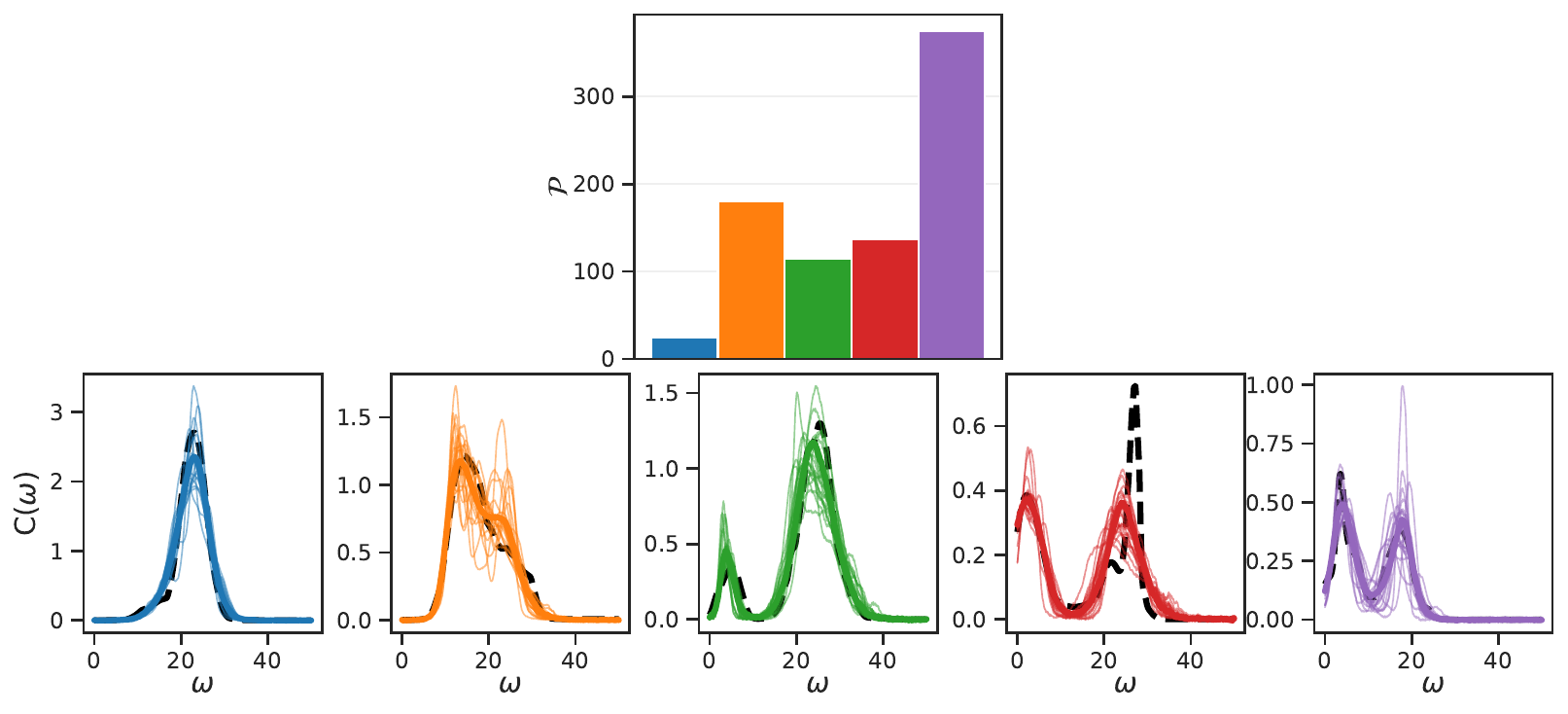}}
\caption{Quantitative analysis of inversion ambiguity for the representative synthetic examples shown in \cref{fig:example_of_ac}. \textbf{Top:} A bar chart displaying the computed \upv{}, $\mathcal{P}$, for each case. \textbf{Bottom:} A visual legend linking the colors in the bar chart to their respective spectral shapes, displaying the corresponding ground-truth (dashed black) and mean reconstructed (solid thick colored lines) spectra, $C(\omega)$, and thin lines represent twenty different realizations with the diffusion model. The \upv{} serves as a reliable diagnostic for the intrinsic difficulty of the inversion: straightforward reconstructions (e.g., the single peak in blue) yield a low \upv{}, reflecting a tightly constrained solution space. As spectral features become more complex and vulnerable to the smoothing of the forward Laplace transform (e.g., the distinct dual peaks in green and red), the \upv{} predictably increases. In inherently difficult inversions where the model struggles to resolve the correct relative physical scales of distinct features (e.g., the purple spectrum), the \upv{} spikes dramatically. This demonstrates that $\mathcal{P}$ can effectively flag ambiguous inversions without requiring access to the ground-truth data.}
\label{fig:ambiguity_metrics}
\end{center}
\end{figure}

In \cref{fig:ambiguity_metrics} we present a practical comparison of $\mathcal{P} \equiv \mathcal{P}_d$, which serves as a proxy for the difficulty of the inversion. We set $\lambda$ to half the value of the maximal $z_i$ across all compared spectra.
For highly ambiguous inversions where the model struggles to resolve the correct physical scale (e.g., the purple spectrum), ${\mathcal{P}}$ is significantly larger than in less ambiguous cases. Consequently, ${\mathcal{P}}$ provides a reliable diagnostic metric to flag ``hard'' inversions that are inherently limited by the integration kernel (e.g. \cref{fig:ill_posed_demo}), all without requiring access to the ground-truth spectrum.  This kind of analysis is clearly valuable: by leveraging diffusion models to learn the conditional distribution and PCA to analyze its structure, we move beyond pointwise uncertainty estimates and recover a compact, interpretable representation of the ambiguity of the inverse problem.

\subsection{Application to Path-Integral Simulation Data}
\begin{figure}
\begin{center}
\centerline{\includegraphics[width=0.35\linewidth]{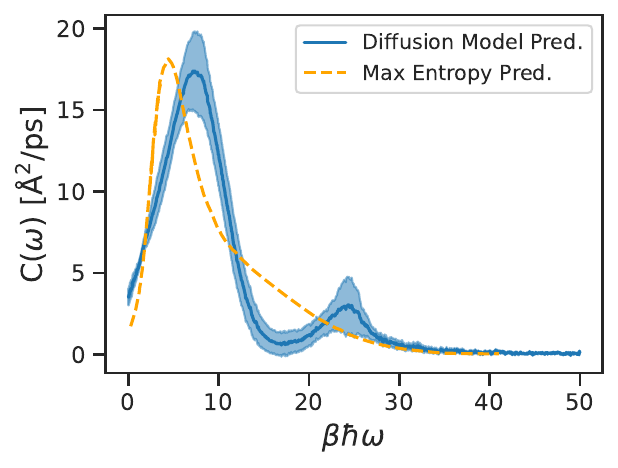}}
\caption{Demonstration of analytic continuation of $G(\tau)$ obtained from PIMD simulation of liquid parahydrogen at $T=14$ K and density $\rho=0.0235$\AA$^{-3}$, compared against maximum entropy solution~\cite{rabani2002calculation}. The diffusion model prediction (solid blue, with shaded uncertainty) generates a high-frequency feature near $\beta\hbar\omega \approx 25$; the wide uncertainty bounds correctly flag this as a region of low confidence, whereas the maximum entropy solution (dashed orange)~\cite{rabani2002calculation} completely smooths it over. The \upv{} is $\mathcal{P} = 61$, indicating moderate inversion ambiguity. Extracted self-diffusion coefficients are $D = 0.28$ \AA$^2$/ps for maximum entropy and $D = 0.59 \pm 0.09$ \AA$^2$/ps for the diffusion model, both comparable to the experimental value of $D = 0.4$ \AA$^2$/ps~\cite{rabani2002calculation}.
}
\label{fig:parahydrogen}
\end{center}
\end{figure}

So far, our analysis has relied on synthetic data, where direct access to the ground-truth spectrum allows an explicit validation. In practical applications, however, this ground truth is unknown. Here, we demonstrate the utility of our framework on noisy, physically realistic data by applying it to an iTCF obtained from a path-integral molecular dynamics (PIMD) simulation of liquid parahydrogen. While standard methods, such as maximum entropy or regression-based NNs, yield single-point estimates without an intrinsic measure of confidence, our diffusion model explicitly quantifies the underlying uncertainty. We simulated the system at 14 K with a density of 0.0235 \AA$^{-3}$, matching the conditions described by \citet{rabani2002calculation}, and compared our generative reconstructions against their maximum entropy solution (\cref{fig:parahydrogen}).

The resulting \upv{} for this inversion is $\mathcal{P} = 61$. Placed in the context of our synthetic analysis (\cref{fig:ambiguity_metrics}), this value indicates a moderate level of intrinsic ambiguity: more challenging than the simplest single-peak spectra, but less ambiguous than the heavily overlapping multimodal cases. Notably, the diffusion model generates a secondary high-frequency peak near $\beta\hbar\omega \approx 25$. Because the forward integral kernel exponentially suppresses higher frequencies, features in this regime are notoriously difficult to reconstruct and are prone to spurious structure~\cite{shi2023rethinking}. However, unlike regression or maximum entropy methods, our diffusion framework captures this ambiguity: the shaded variance around this secondary peak is substantial, indicating lower confidence in this high-frequency structure. 

Furthermore, we evaluate the self-diffusion coefficient, $D$, extracting it from the zero-frequency limit of the spectrum. The maximum entropy reconstruction yields $D=0.28$ \AA$^2$/ps, while our diffusion model predicts $D=0.59\pm0.09$ \AA$^2$/ps. Both computational methods provide estimates broadly comparable to the experimental value of $D=0.4$ \AA$^2$/ps~\cite{rabani2002calculation}. However, our diffusion framework uniquely provides a principled uncertainty bound derived directly from the learned posterior distribution. We stress that this application is a demonstration rather than a benchmark. No exact reference spectrum exists for liquid parahydrogen, so neither reconstruction can be scored against ground truth. The underlying ambiguity is common to all analytic continuation methods, which usually leave it implicit. What our framework adds is a way to show which spectral features the data support and which they do not, provided the model was trained on spectra representative of the system at hand.

\section{Summary and Conclusion}

Analytic continuation of iTCFs is a notoriously ill-posed inverse problem. The inversion is fundamentally non-unique. Multiple physically distinct spectra can yield nearly identical imaginary-time observations (see \cref{fig:ill_posed_demo}). Traditional regularization techniques, such as maximum entropy, and recent regression-based machine learning models mask this ambiguity by collapsing the solution space towards the conditional mean. To overcome this limitation, we introduced a diffusion-based generative framework for numerical analytic continuation. Instead of forcing a deterministic one-to-one mapping, our approach models the full conditional probability distribution of admissible spectra, $p(C \mid G)$. This probabilistic framing preserves the intrinsic multiplicity of the solution space and provides a theoretically grounded basis for uncertainty quantification directly from the generated ensembles. Furthermore, through a PCA-based analysis, we demonstrated that the uncertainty in the reconstructed spectra is highly correlated across frequencies, revealing that the physically meaningful ambiguity is not random noise but is confined to a low-dimensional manifold.

Building on this structural analysis, we introduced the \upv{}, a metric that characterizes the magnitude of this correlated uncertainty. We showed that the \upv{} serves as a reliable diagnostic tool to assess the intrinsic ``hardness'' of an inversion task, successfully flagging highly ambiguous reconstructions without requiring access to the ground-truth spectrum. We validated our framework on PIMD simulation data for liquid parahydrogen, where our model extracted transport properties with an uncertainty bound and identified high-frequency spectral regions where the inverse problem is fundamentally unconstrained.
Finally, quantifying what a model does not know advances on classical and regression-based methods. By explicitly identifying which spectral features are reliable and which are artifacts of the ill-posed kernel, our diffusion framework provides a more trustworthy, data-driven tool for bridging imaginary-time simulations and real-time observables.

\section{Computational details}
\label{sec:comp_details}

\subsection{Model Architecture} 
The generative model employs a 1D Diffusion Transformer (DiT) architecture tailored for continuous-time diffusion. The model processes discretized real-frequency spectra, $C(\omega)$, which are represented as 1D arrays of 1024 points. The input spectra are first patchified using a 1D convolutional layer with a patch size and stride of 2. The core network comprises 12 Transformer decoder blocks with a latent embedding dimension of 256. Each block utilizes non-causal self-attention, cross-attention for conditioning, and a multilayer perceptron (MLP). The attention mechanisms employ 4 heads. To enhance local structural inductive biases, the standard MLP is replaced with a depthwise separable 1D convolution block (inspired by LocalViT~\cite{li2021localvit}) using an expansion multiplier of 4. Dropout was set to zero throughout the network. 

For fair comparison, the deterministic regression baseline utilized the identical 12-layer DiT core architecture. To adapt the network for direct regression without a noisy spectral input, the input pathway was bifurcated: first, a dedicated projection network consisting of linear transformations and a strided 1D convolution compresses and expands the 4-channel, 99-point $G(\tau)$ input into a 1-channel, 1024-point surrogate sequence. This sequence is then fed into the core network, where it is patchified and processed identically to a spectral input. Simultaneously, the original 4-channel $G(\tau)$ signal is linearly projected and fed into the cross-attention layers. This ensures the regression baseline leverages the exact same structural capacity, patchification, and conditioning pathways as the diffusion model.

\subsection{Data Generation} 
Synthetic training data was generated procedurally on-the-fly, ensuring the model was exposed to a continuous stream of novel spectra without repeating a fixed dataset. The real-frequency spectra were constructed as random mixtures of 1 to 4 structural bumps defined over a frequency domain of $\omega \in (0, 50)$. To mimic realistic, asymmetrical physical spectra, these bumps were not simple Gaussians; instead, they were procedurally warped using spline functions on the $\omega$ domain, where the splines were generated using a stochastic sequence of 5 to 50 control points governed by randomized relative inertia parameters. The individual bumps were then smoothed and normalized. To ensure a diverse distribution of spectral features, the center of each bump was uniformly sampled within the lower half of the frequency domain (0\% to 50\% of the maximum frequency, corresponding to $\omega \in [0, 25]$). The widths of these bumps were also randomly drawn to span between 10\% and 60\% of the total domain length. 

Once the random bumps were linearly combined into a single raw spectrum, a rigorous multi-stage normalization protocol was applied to ensure physical and numerical consistency across the dataset. First, the combined spectrum was scaled by its maximum value so that the peak intensity of the raw spectrum equaled 1. Next, the corresponding imaginary-time correlation function, $G(\tau)$, was computed over a 99-point grid by numerically evaluating the forward Laplace integral transform using the trapezoidal rule. Finally, to ensure strict area-normalization, the total integrated area under the computed $G(\tau)$ curve was evaluated via a second trapezoidal integration. Both the real-frequency spectrum $C(\omega)$ and the iTCF $G(\tau)$ were then divided by this area. This global normalization guarantees that the integral of $G(\tau)$ over the temporal grid is always exactly equal to $1/\pi$, providing a scale-invariant optimization target for the diffusion model.

\subsection{Training Hyperparameters} 
Due to the on-the-fly generation of the dataset, traditional epoch counting reflects the generation of discrete batches of novel samples rather than multiple passes over a fixed dataset. The model was trained for 800 nominal epochs, with each epoch consisting of 20,000 dynamically generated spectra. Using a batch size of 50, this equates to 320,000 total gradient steps, exposing the model to 16 million unique, procedurally generated training samples throughout the optimization process. The network was optimized using the Adam optimizer with a learning rate of $3 \times 10^{-5}$. The loss function evaluated the Mean Squared Error (MSE) between the predicted and ground-truth clean spectra. To accelerate training, FP16 mixed-precision was employed. During training, the continuous diffusion time $t$ was sampled from a Beta distribution parameterized by $\alpha=1$ and $\beta=2.5$. This specific distribution skews the sampled timesteps toward zero, deliberately placing greater emphasis on minimizing the loss in low-noise regimes where the fine structural refinement of the spectra occurs. Furthermore, to enable classifier-free guidance during inference, 15\% of the conditioning $G(\tau)$ labels were randomly masked (set to zero) during the training loops. Finally, an Exponential Moving Average (EMA) of the model weights was maintained with a decay factor of 0.999; this smoothed EMA model was used for all final evaluations and inferences.

\subsection{Data availability}
The data that support the findings of this study are openly available at:

\url{https://github.com/Hirshberg-Lab/transformer_latent_diffusion_4AC}

\section*{Acknowledgments}
B.H. acknowledges support by the Israel Science Foundation (grants No.\ 1037/22 and 1312/22). S.M. acknowledges support from the Tel Aviv University Center for Artificial Intelligence and Data Science (TAD) and the Israeli Planning and Budgeting Committee (VATAT).

\bibliography{refs}

\newpage

\renewcommand{\thefigure}{S\arabic{figure}}
\renewcommand{\thetable}{S\arabic{table}}
\renewcommand{\theequation}{S\arabic{equation}}
\renewcommand{\thesection}{S\arabic{section}}
\setcounter{figure}{0}
\setcounter{table}{0}
\setcounter{equation}{0}
\setcounter{section}{0}

\section*{Supporting Information}

\section{Conditioning Strategy}
\label{si:conditioning}
To provide a richer conditional representation and expose hidden structural features, the raw $G(\tau)$ signal was transformed into four complementary channels within the data loader before entering the network. These parallel channels consist of: (1) the raw $G(\tau)$ signal, (2) the natural logarithm of the signal, (3) a pointwise normalized signal derived from precomputed dataset means and absolute deviations, and (4) a pointwise empirical Cumulative Distribution Function (CDF) mapping. These four channels were then standardized using global channel-wise means and standard deviations. The continuous diffusion time $t$ was encoded using a sinusoidal embedding. This noise embedding was concatenated with a linearly projected representation of the four-channel $G(\tau)$ label. The resulting combined conditioning embedding was subsequently fed into the network via the cross-attention layers of each Transformer block.

\section{Sampling Settings}
\label{si:inference}
To generate real-frequency spectra from the trained generative model, the reverse diffusion process was initialized from pure standard Gaussian noise, $x_{t_0} \sim \mathcal{N}(0, \mathbb{I})$. The inference was performed over a discrete, linearly spaced sequence of $N=40$ diffusion timesteps, starting from an initial noise level of $t_1=0.99$ and decreasing progressively to $t_N \approx 0$.
To accelerate convergence while maintaining high sample quality, we employed a hybrid solver based on DPM-Solver++~\cite{lu2025dpm}. Specifically, the first denoising iteration utilized a standard first-order DDIM step. For the remaining 39 steps, the algorithm leveraged a second-order DPM-Solver++ update, which computes a higher-order effective prediction by extrapolating between the current and previous clean spectrum estimates ($\hat{x}_0^{(i)}$ and $\hat{x}_0^{(i-1)}$).

Although the network was trained with 15\% label dropout to support classifier-free guidance, the final evaluations and reconstructions were performed using pure conditional sampling (guidance scale set to $1.0$). This ensures the generated samples faithfully represent the unbiased conditional posterior distribution $p(C \mid G)$ without artificial structural sharpening. Finally, to construct the uncertainty bounds, an ensemble of 1000 independent spectral realizations was generated for each test iTCF by repeatedly sampling different initial noise vectors.

\section{Algorithms}
\label{si:algorithms}
\Cref{alg:training} and \cref{alg:inference} summarize the training and sampling procedures, respectively.

\begin{algorithm}[H]
\caption{Training the Diffusion model}\label{alg:training}
\KwData{Dataset of ground-truth spectra and corresponding iTCFs $\{ (x_0, G) \}$}
\KwResult{Trained network parameters $\theta$}
 Initialize network parameters $\theta$\;
 \While{not converged}{
  Sample a batch of pairs $(x_0, G)$ from the dataset\;
  Sample timesteps $t \sim \text{Beta}(\alpha, \beta)$ \Comment*[r]{Skewed towards $0$}
  Sample noise $\epsilon \sim \mathcal{N}(0, \mathbb{I})$\;
  $x_t \gets (1-t)x_0 + t\epsilon$ \Comment*[r]{Continuous-time forward process}
  $\hat{x}_0 \gets x_\theta(x_t, t, G)$ \Comment*[r]{Predict clean spectra}
  $L(\theta) \gets \| \hat{x}_0 - x_0 \|^2$ \Comment*[r]{Compute loss}
  $\nabla_\theta L(\theta)$ and update $\theta$ \Comment*[r]{Backpropagate and update}
 }
\end{algorithm}

\begin{algorithm}[H]
\caption{Inference via DPM-Solver++}\label{alg:inference}
\KwData{Input $G$, number of steps $N$, sequence $\{t_i\}_{i=1}^N$, sequence $h_{i} = \ln((1-t_{i})/t_{i})-\ln((1-t_{i-1})/t_{i-1})$} 
\KwResult{Reconstructed spectral sample $x_0$}
 $x_{t} \sim \mathcal{N}(0, \mathbb{I})$ \Comment*[r]{Initialize pure noise}
 \For{$i \gets 1$ \KwTo $N-1$}{
  $\hat{x}_0 \gets x_\theta(x_{t}, t_i, G)$ \Comment*[r]{Predict denoised spectra}
  
  \eIf{$i = 1$}{
   $\hat{D} \gets \hat{x}_0$ \Comment*[r]{First-order DDIM step}
  }{
   $\hat{D} \gets \left(1 + \frac{h_{i+1}}{2 h_{i}}\right) \hat{x}_0 - \frac{h_{i+1}}{2 h_{i}} \hat{x}_{0\text{,previous}}$ \Comment*[r]{Higher-order prediction}
  }
  $x_{t} \gets  \left( (t_i - t_{i+1}) \hat{D} + t_{i+1} x_{t} \right)/t_i$ \Comment*[r]{Deterministic reverse step}
  $\hat{x}_{0\text{,previous}} \gets \hat{x}_0$\;
 }
 \Return{$x_\theta(x_{t}, t_N, G)$} 
\end{algorithm}

\section{Principal Components of the Diffusion Ensemble}
\label{si:pca}
\begin{figure}[H]
\begin{center}
\centerline{\includegraphics[width=0.79\linewidth]{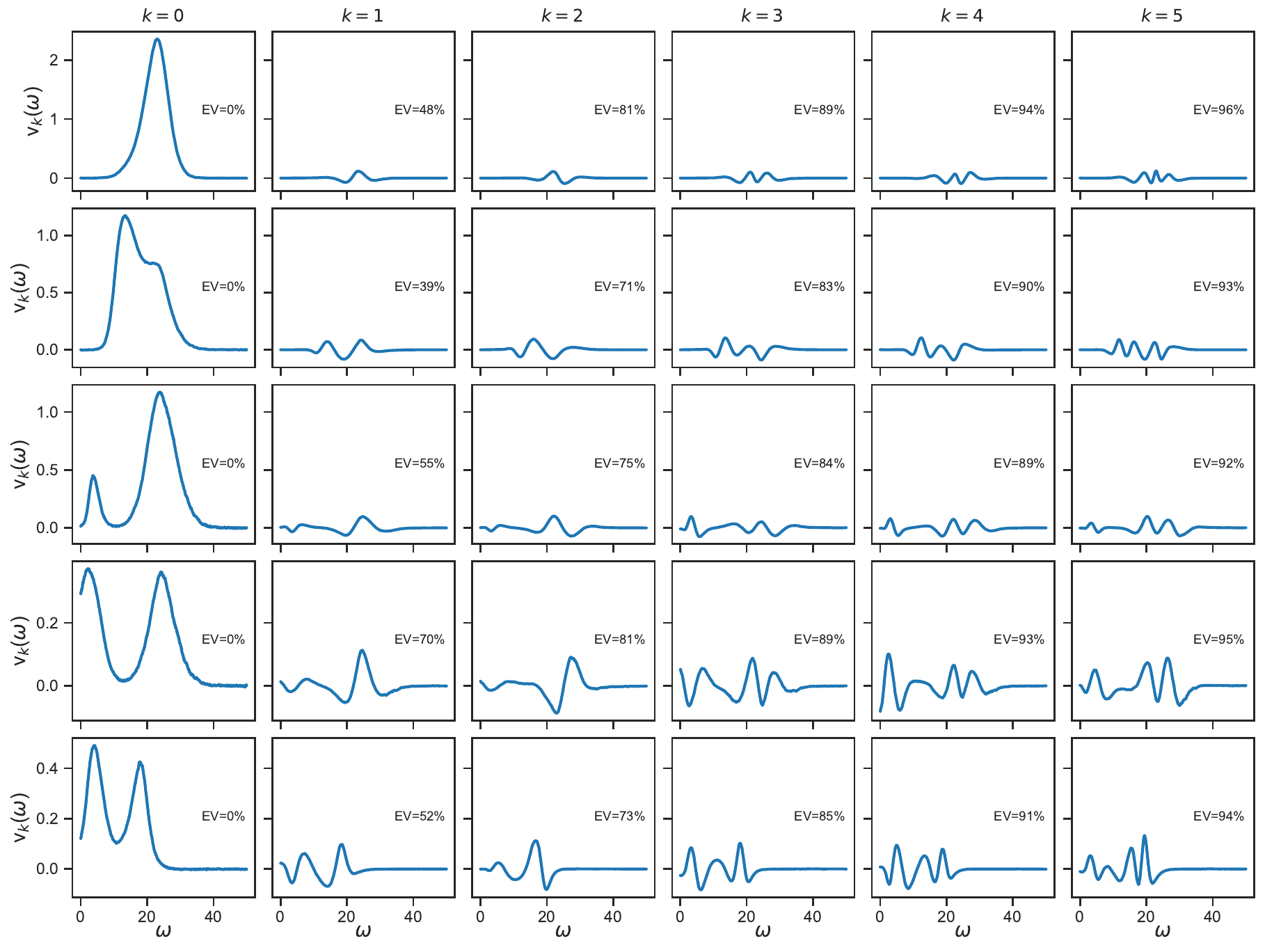}}
\caption{Principal components of the diffusion ensemble for five representative iTCFs (rows). The first column ($k=0$) is the ensemble mean $\mu$, and the remaining columns show the first five principal components $v_k(\omega)$, which become increasingly oscillatory with $k$. Annotations give the cumulative explained variance (EV), which exceeds $90\%$ by $k=5$ in every case.}
\label{fig:vis_pca_comp}
\end{center}
\end{figure}

\Cref{fig:vis_pca_comp} shows the leading principal components of the diffusion ensemble for five representative iTCFs.

\end{document}